\documentclass[useAMS,usenatbib]{mn2e}

\usepackage{color}
\usepackage{graphicx}
\usepackage{bbold}
\usepackage{subfigure}
\usepackage{breqn}

\title[Circumplanetary disk or circumplanetary envelope?]{Circumplanetary disk or circumplanetary envelope?}
\author[J. Szul\'agyi et al.]{J. Szul\'agyi$^{1,2}$\thanks{E-mail:
judits@phys.ethz.ch} F. Masset$^{3}$, E. Lega$^{1}$, A. Crida$^{1,4}$, A. Morbidelli$^{1}$, and T. Guillot$^{1}$\\
$^{1}$Laboratoire Lagrange, Universit\'e C\^ote d'Azur, Observatoire de la C\^ote d'Azur, CNRS,
Blvd de l'Observatoire, CS 34229,\\ 06304 Nice cedex 4, France\\
$^{2}$ ETH Z\"urich, Institute for Astronomy, Wolfgang-Pauli-Strasse 27, CH-8093, Z\"urich, Switzerland\\
$^{3}$Instituto de Ciencias F\'\i sicas, Universidad Nacional Aut\'onoma de M\'exico, Avenidad Universidad s/n, 62210 Cuernavaca,\\ Morelos, Mexico\\
$^{4}$Institut Universitaire de France, 103 Bd. Saint-Michel, 75005 Paris, France}
\begin{document}

\date{Accepted XX. Received XX; in original form 2015 September 09}

\pagerange{\pageref{firstpage}--\pageref{lastpage}} \pubyear{2015}

\maketitle

\label{firstpage}

\begin{abstract}
We present three-dimensional simulations with nested meshes of the dynamics of the gas around a Jupiter mass planet with the JUPITER and FARGOCA codes. We implemented a radiative transfer module into the JUPITER code to account for realistic heating and cooling of the gas. We focus on the circumplanetary gas flow, determining its characteristics at very high resolution ($80\%$ of Jupiter's diameter).  In our nominal simulation where the temperature evolves freely by the radiative module and reaches 13000\,K at the planet, a circumplanetary envelope was formed filling the entire Roche-lobe. Because of our equation of state is simplified and probably overestimates the temperature, we also performed simulations with limited maximal temperatures in the planet region (1000\,K, 1500\,K, and 2000\,K). In these fixed temperature cases circumplanetary disks (CPDs) were formed. This suggests that the capability to form a circumplanetary disk is not simply linked to the mass of the planet and its ability to open a gap. Instead, the   gas temperature at the planet's location, which depends on its accretion history, plays  also fundamental role. The CPDs in the simulations are hot and cooling very slowly, they have very steep temperature and density profiles, and are strongly sub-Keplerian. Moreover, the CPDs are fed by a strong vertical influx, which shocks on the CPD surfaces creating a hot and luminous shock-front. In contrast, the pressure supported circumplanetary envelope is characterized by internal convection and almost stalled rotation.
\end{abstract}


\section{Introduction}

The importance of studying the circumplanetary disk (CPD) formed around massive giant planets is twofold: this subdisk regulates the growth of the planet in the last stages (e.g. \citealt{Lissauer09, Rivier12, Szulagyi14}), and it is the birth-nest for satellites to form  \citep{CW02,CW06,ME03a,ME03b}. Currently there is no unambiguous detection of CPD from observations, although extended thermal emission was detected around the planetary candidates of LkCa15 \citep{KI12}, HD100546 \citep{Quanz13,Quanz14}, HD 169142 \citep{Reggiani14}, and the upper limit of CPD masses were measured in the system of GSC 6214-210 \citep{Bowler15} with the Atacama Large Millimeter Array. Until sufficiently resolved CPDs are observed, hydrodynamic simulations of these subdisks are the only tool to understand and reveal their characteristics. 
As computers evolve, more and more complex (and accurate) physical models are used in the hydrodynamic simulations as well. But resolving well the circumplanetary disk is challenging even in numerical simulations. One way to do simulations of the CPD is to perform 2D calculations    (such as \citealt{Rivier12, DA03, Lubow99,KD06}), where sufficiently high resolution can be achievable due to the limitations on 2 spatial directions. However, e.g.  \citet{Bate03,Tanigawa12, Morbidelli14, Szulagyi14,PM08,Gressel13} showed that the third dimension really changes the whole picture on the accretion flow of gas to the planet, thus also on the role of the subdisk, regardless the equation-of-state used. In three-dimensional simulations, another possible way to resolve the CPD is to limit the simulation box size. Instead of simulating the entire circumstellar disk, one can define a box in the vicinity of the planet, where the simulation is performed. These are the so-called shearing sheet box simulations, such as \citet{Machida10, Tanigawa12}. However, this way the planetary gap is not deep enough, and the CPD is missing the feedback from the circumstellar disk. In \citet{Morbidelli14} and in \citet{Szulagyi14} it is described that the accretion onto the CPD is a free-fall flow arising from the top layers of the circumstellar disk, which is part of an entire meridional circulation flow between the CPD and the circumstellar disk. With a shearing sheet box simulation, this meridional circulation is missed, therefore the accretion and the CPD dynamics are not correctly addressed. Hence, our approach in this paper is to do a global disk simulation featuring the entire circumstellar disk, and using so-called nested meshes, i.e. high resolution grids around the planet, to zoom onto the planet's vicinity. We also added a complex radiative module to our JUPITER code, incorporating the thermal processes (viscous heating, stellar irradiation, and cooling through radiation), assuming a uniform dust-to-gas ratio. The JUPITER code is based on a shock capturing method, therefore sharp details can be observed in the flow around the planet, which were never shown before. 

Previous non-isothermal simulations addressing the subdisk (e.g. \citealt{AB09a,AB09b,DA03,DA15,Gressel13}) have already pointed out that the CPD is hot, its inner part is optically thick, and it has a steep radial temperature profile. Works of \citet{DA03}, \citet{AB09a}, and \citet{Gressel13} also agree that, due to the high temperatures, the spiral density wake in the CPD is less prominent than in isothermal simulations, suggesting a reduced stellar torque,    that can alter accretion. To characterize the CPD,    apart from a proper thermal model, it is also important what resolution the simulations use and what is the applied smoothing technique for the gravitational potential of the planet, since these factors highly affect the resulting temperatures. With our new code, we examine here the circumplanetary material around a 1 $\mathrm{M_{Jup}}$ planet with maximal resolution of $\sim 80\%$ of Jupiter-diameter, but with smoothening of the planetary potential which accounts for almost five Jupiter-diameter. We also compare our results with a previously existing hydro-code with radiative module, called FARGOCA \citep{FARGOCA}.

\section[]{Physical Model and Numerical Methods}
\label{sec:numerical}

Our study is based on three-dimensional, grid-based hydrodynamic simulations with the JUPITER code \citep{Szulagyi14,Borro06}, originally developed by F. Masset. This code is based on a higher order Godunov scheme and has nested meshes, which allows to zoom onto the planet's vicinity with high resolution. The original code included only a locally isothermal equation of state (EOS); here we implemented an energy equation (so called adiabatic EOS) and a simplified radiative transfer module to account for realistic heating and cooling. The radiative module follows a two-temperature approach in the grey approximation (one group of photons) and uses Fluxes Limited Diffusion (FLD), according to the governing equations of \citet{Bitsch13} and \citet{Bitsch14}. The multi-level strategy for the radiative transfer is that of \citet{Commercon11}, using Dirichlet boundary conditions for the coarse-to-fine and fine-to-coarse interfaces. Apart from the equations of the mass  and momenta conservation, the code now solves the energy equation on the total energy, and accounts for the coupling between the thermal energy and radiation energy ($\epsilon_{rad}$). The full set of governing equations reads therefore:

\begin{equation}
{\partial \rho \over \partial t}+\nabla \cdot (\rho \mathbf{v}) = 0
\label{eq:cont}
\end{equation}

\begin{equation}
{\partial (\rho \mathbf{v}) \over \partial t}+\nabla \cdot (\rho \mathbf{v} \otimes \mathbf{v})+\nabla {P} = -\rho \nabla{\Phi} + \nabla \cdot {\bar{\bar{\tau}}}
\label{eq:mom}
\end{equation}

\begin{equation}
{\partial E \over \partial t}+\nabla \cdot \left[\left(P{\bar{\bar{ \mathrm{I}}}}-{\bar{\bar{\tau}}}\right)\cdot\mathbf{v} + E\mathbf{v}  \right]  = \rho \mathbf{v} \cdot \nabla{\Phi} - \rho \kappa_P c \left[ \frac{B(T)}{c} - \epsilon_{rad}\right]
\label{eq:ene}
\end{equation}

\begin{equation}
{\partial \epsilon_{rad} \over \partial t}=-\nabla \cdot F_{rad} +\rho \kappa_P c \left[ \frac{B(T)}{c} - \epsilon_{rad}\right]
\label{eq:erad}
\end{equation}

where $\rho$ is the gas density, $E$ is the gas total energy (sum of the internal and kinetic energies), $\mathbf{v}$ stands for the velocity-vector, and $P$ indicates the pressure. Moreover, $\Phi$ is the gravitational potential, $c$ indicates the speed of light, and $B(T)$ defines the thermal blackbody: $ 4 \sigma T^4$ -- here $\sigma$ symbolizes the Stefan-Boltzmann constant, while $T$ stands for the temperature. The Planck mean opacity, $\kappa_P$, is defined as Eq. 13 in \citet{Bitsch13}. Eq. \ref{eq:ene} contains the unit matrix (${\bar{\bar{\mathrm{I}}}}$) and the stress-tensor (${\bar{\bar{\tau}}}$), which is defined as:

\begin{equation}
\bar{ \bar \tau} = 2\rho \nu \left[\bar{ \bar  \mathrm{D}}-{1\over 3}\left(\nabla \cdot \mathbf{v}\right){\bar{\bar{ \mathrm{I}}}}\right]
\label{st}
\end{equation}

where $\nu$ is the kinematic viscosity, and $\bar{ \bar  \mathrm{D}}$ is the strain tensor. Furthermore, $F_{rad}$ in Eq. \ref{eq:erad} is defined as:

\begin{equation}
F_{rad}=-\frac{c\lambda}{\rho \kappa_R} \nabla \epsilon_{rad}
\label{eq:frad}
\end{equation}

where $\lambda$ is the flux limiter taking care of the smooth transition between optically thin and thick domains. For its definition and usage, see \citet{Kley89} and \citet{Bitsch13}. The parameter $\kappa_R$ indicates the Rosseland mean opacity, which is defined as Eq. 15 in \citet{Bitsch13} and chosen to be equal to Planck mean opacity. We assume a uniform dust to gas ratio (generally 0.01 but this can be modified as an input parameter of the code), and use the \citet{BL94} opacity tables. 

To close the system of equations, the EOS also needs to be defined. We used an adiabatic EOS with the adiabatic index ($\gamma$) equal to 1.43:
\begin{eqnarray}
P=(\gamma-1)\epsilon
\end{eqnarray}
where $\epsilon$ is the internal energy of the gas, which is $\epsilon=\rho c_v T$. 
The different physical parts corresponding to the governing equations (hydrodynamics and radiative transfer) are solved in succession using operator splitting. Namely, a full update of the quantities over a timestep consists of the following substeps:
\begin{enumerate}
\item We solve for the interface fluxes (Eqs. \ref{eq:cont}--\ref{eq:ene}) using an exact adiabatic Riemann solver, for which the left and right states are obtained using the spatially second order accurate scheme of MUSCL-Hancock \citep[predictor step]{Toro_book}.
\item These fluxes are used to update the cell contents, in what is called a ``conservative update" (because the fluxes are shared on interfaces between adjacent cells, the scheme is conservative
for the corresponding quantities to platform accuracy).  Prior to being used, the momenta and energy fluxes are corrected (augmented) by the viscous stresses.
\item Source terms (gravitational and fictitious forces) are applied using a finite difference scheme.
\item In the radiative module, Eq. \ref{eq:erad} is solved and the internal energy is updated through the two-temperature approach explained in \citet{Bitsch13}.
\end{enumerate}

The primitive variables in the JUPITER code are the volume density, the three components of the velocity, and the internal energy. In each cell in the simulation the values of density/energy/velocity are all centered values, and thus defined at the coordinates of the cell barycenters.

The new parts of the JUPITER code (energy equation, radiative module) were heavily tested both separately and together with the hydro-kernel as well. This includes testing the adiabatic Riemann solver, the inter-CPU and inter-level communications, testing the radiative module against other hydro codes with radiative transfer, and checking the dimensional homogeneity of all the equations in the entire code. For details of the testing, see \citep{Szulagyi_phd}.

For comparison purposes we carried out a simulation with the FARGOCA code \citep{FARGOCA} as well, which is an improvement of the FARGO code  \citep{Masset00} in 3D. FARGOCA is finite difference, staggered mesh code based on an upwind method with van Leer's slopes. It solves the energy equation directly for the internal energy: 

\begin{equation}
{\partial \epsilon \over \partial t}+\nabla \cdot (\epsilon \mathbf{v}) = Q^+ -P \nabla \cdot \mathbf{v} -  \rho \kappa_P c \left[ \frac{B(T)}{c} - \epsilon_{rad}\right]
\label{eq:enefargo}
\end{equation}

where $ Q^+ = ({\bar{\bar{\tau}}} \nabla) \cdot \mathbf{v}$ is the viscous heating. While JUPITER is based on a shock-capturing method, FARGOCA solves the hydrodynamic equations explicitly, thus it is not as well suited as JUPITER to treat shocks. On the other hand, treating directly the equation for the internal energy alleviates the so-called high Mach number problem \citep{Ryu93,TP04} faced in JUPITER, where the internal energy is very small fraction of the total energy. The FARGOCA code does not have nested meshes that are needed to reach the same high resolution around the planet as with JUPITER. Therefore, we have implemented a manual  refinement procedure in FARGOCA as follows. First, for the gap-opening phase the global disk simulation was performed on the same resolution as JUPITER's coarsest mesh. Then, a box around the planet was defined, where the resolution was doubled, and the boundary conditions for all hydrodynamics variables were set to the values found on the coarser mesh. After the gas flow has stabilized on a given resolution, the box size around the planet was reduced again, the resolution was doubled and the same boundary procedure was adopted.
 This procedure was iterated until we achieved a simulation box of size slighly larger than the gap. This is the minumum box size which allows to properly take into account both the gap and the meridional circulation with the circumstellar disk.  In order to run simulations on a resonable CPU time \footnote{45  days on 160 processors for 60 Jupiter's orbits on about 82 million grid cells.} we have used on this box a resolution which is   half of the resolution in JUPITER's finest level.

\section[]{Setup of the Simulations}
\label{sec:setup}

\subsection{Units, Frame and Grid}

In our simulations, the coordinate system is spherical -- with coordinates of azimuth, radius, co-latitude -- centered onto the star. The mass unit is the mass of the central star (assumed to be solar in numerical applications), the length unit is the radius of the planetary orbit ($r_p$), while the time unit is such that $G$, the gravitational constant, is one. This implies that the planetary orbital period, $2\pi /\Omega=2\pi/\sqrt{G(M_*+M_p)/r_p^3}$, is $2 \pi$. The frame is co-rotating with the planet, so that the planet is at a fixed position throughout the simulation. This position is at azimuth=0.0, radius=1.0, co-latitude=$\pi/2$. Assuming the planet's orbit is at Jupiter's distance, the length unit in the code is 5.2\,AU. The radial limits of the simulation box are 0.4-2.4 code units (2.1-12.5 AU), the azimuthal range is from $-\pi$ to $+\pi$, thus including the entire circumstellar disk, and the co-latitude range is [1.442, $\pi/2$], with the mid-plane being on $\pi/2$. This means a 7.4 degree opening angle for the circumstellar disk. To save computational time, we simulated only the half of the circumstellar disk, thus assuming symmetry relative to the midplane. 

Due to the nested meshes, the resolution changes grid level by grid level. On the coarsest mesh (level 0), the resolution was $680\times215\times20$, which means $dr=0.009$\,code units$=0.048\,$AU. On the next level, the resolution was double of the previous, and so on, till level 6. In other words, at each level refinement the resolution doubled in each spatial direction. The highest level of resolution (reached on level 6) was $dr=1.442\times10^{-4}$code units\,$=7.498\times10^{-4}\,$AU$\,=\,0.8\,\mathrm{d_{Jup}}$ where $\mathrm{d_{Jup}}$ is the diameter of Jupiter. The borders of the nested meshes are described in Table \ref{tab::meshing}. The simulation began with only the coarsest mesh, and the successive levels were added in sequence after a quasi-steady state was reached on the previous level.

\begin{table*}
  \caption{Number of cells on different grid levels}
 \label{tab::meshing}
  \begin{tabular}{p{.7cm}p{1.2cm}p{1cm}p{1.2cm}p{3.7cm}p{3.5cm}p{2.5cm}}
  \hline
   Level    &  $N^{\circ}$ of cells in azimuth  &  $N^{\circ}$ of cells in radius &  $N^{\circ}$ of cells in co-latitude & Boundaries of the levels in azimuth [rad] &  Boundaries of the levels in radius [a] & Boundaries of the levels in co-latitude [rad]\\
 \hline
       0	&	680	&	215	&	20 & [$-\pi$, $\pi$] & [0.40005, 2.3845] & [1.4416, $\pi/2$]\\
       1	&	120	& 	120	& 	34 & [-0.27735, 0.27735] & [0.72264, 1.27735] & [1.451041, $\pi/2$]\\
       2	&	120	&	120	&	62 & [-0.138675, 0.138675] & [0.861325, 1.138675] & [1.47137, $\pi/2$]\\
       3	&	120	& 	120	&	86 & [-0.0693375, 0.0693375] & [0.9306625, 1.0693375] & [1.5014588, $\pi/2$]\\
       4	&	120	&	120	& 	86 & [-0.03466875, 0.03466875]& [0.96533125,1.03468875] & [1.5361276, $\pi/2$]\\
       5	&	120	&	120	& 	86 & [-0.017334375, 0.017334375] & [0.98266562, 1.017334375] & [1.553462, $\pi/2$]\\
       6	&	120	&	120	& 	86 & [-0.0086671875, 0.00866719] & [0.99133281, 1.0086671875] & [1.5621291, $\pi/2$]\\
       7	&	120	&	120	& 	86 & [-0.0043335938, 0.00433359] & [0.99566641, 1.0043335938] & [1.566462737, $\pi/2$]\\
\end{tabular}
\end{table*}

\subsection{Boundary Conditions}

At the radial boundaries of the simulation box, we have used reflecting boundary conditions for the radial velocity. The azimuthal velocity was extrapolated in the 2 ghost cells according to the local Keplerian velocity. The density and energy values in the ghost cells were set equal to the corresponding values of their images among the active zones.

At the midplane (colatitude = $\pi/2$), a reflecting boundary condition is applied. At the other edge in colatitude, above the circumstellar disk surface layer, we fix the temperature to 30 Kelvin in the ghost cells to force the cooling. This accounts for the fact that circumstellar disks are surrounded by the outer space and are able to radiate away their heat. In the azimuthal direction, periodic boundary conditions were used. 

At the border between nested meshes, the flow should be smooth, therefore the JUPITER code uses a complex ghost cell communication with multi-linear interpolation. This means that, in each direction, 2 cells overlap with the cells of the previous level beyond the border of the given mesh, and in these ghost cells the hydrodynamic fields are linearly interpolated from the values available on the previous level.

\subsection{Disk Physics}

The circumstellar disk's initial surface density was $\Sigma=\Sigma_0(\frac{r}{a})^{-0.5}$ with $\Sigma_0=6.76 \times 10^{-4}$ code units. This density was chosen to be close to the Minimum Mass Solar Nebula (MMSN; \citealt{Hayashi}). The initial disk aspect-ratio was chosen to be $H/r=0.05$, where $H$ is the pressure scale-height of the disk, but this changes as the circumstellar disk cools and therefore contracts  a bit towards the midplane. All of our simulations have a constant viscosity with value $10^{-5} a^2\Omega_p$, which corresponds to approximately a value of $\alpha$ of 0.004 at Jupiter's orbit in the representation of \citet{sunayev}. We remind the reader that, like every numerical simulation, ours are also affected by the numerical viscosity, in particular close to the planet where the mesh is locally Cartesian, while the flow has rather a  cylindrical symmetry. 

Because the opacity table accounts for the dust as well, one needs to define the dust-to-gas ratio in the simulations. We used the interstellar medium value of 1\% dust. The cooling happens through radiation, therefore one should allow energy to escape through the surface of the circumstellar disk. To minimize the CPU-time required to reach the initial thermal equilibrium of the circumstellar disk, we run initially the simulation only with the circumstellar disk (i.e. without a planet). Since the circumstellar disk is azimuthally symmetric, we defined only 2 cells in azimuth, run the simulation until thermal equilibrium was reached, we divided the 2 azimuthal cells into the final amount of 680 cells. From this point on we started to build up the planet increasing its mass over 30 orbits (see below), and run the simulation for another 120 orbits to reach an equilibrium after the gap opening. Only after this we started to add the nested meshes. 

\subsection{Planetary Potential}

To allow the gas flow to adapt to the presence of our heavy planet, we increased its mass gradually over 30 orbit as 
\begin{eqnarray}
M_{p}(t) = M_{p_{\mathrm{final}}} \sin^2 \left(\frac{t}{120} \right)
\end{eqnarray}
This meant that the final planet mass was reached after the first 30 orbits of the simulation time. In all our simulations, $M_{p_{\mathrm{final}}}$ is $10^{-3}$ code unit, which corresponds to a Jupiter mass planet around a solar mass star.

In the simulations the planet is a point-mass, in the corner of 8 cells (of which only 4 are considered as active cells and 4 ghost cells  due to the symmetry relative to the disk's midplane). This means that there is only a potential-well and no physical sphere is modeled for the planet. Hence, no boundary condition is needed around the planet. However, to avoid the singularity of the gravitational potential, we applied the traditional smoothing of the potential on a length $r_s$:
\begin{eqnarray}
U_p&=&-\frac{G M_p}{\sqrt{x_d^2+y_d^2+z_d^2+{r_s}^2}}
\end{eqnarray}
where $x_d=x-x_p$, $y_d=y-y_p$, and $z_d=z-z_p$ are the distance-vector components from the planet in Cartesian coordinates. Our smoothing length $r_s$ was set equal to three times of the cell diagonal on levels 0-5, and 6 cell diagonals on level 6. In other words the potential well was not deepened on level 6 relative to the previous value on level 5. Because the smoothing length changes on every level to avoid the harsh transition of smoothing length when adding a new level, we gradually reduced the smoothing length as $r_{s}(t) = 0.5\left(r_{s_{\mathrm{previous}}} \cos^2 \left(\frac{t-t_0}{4} \right)+r_{s_{\mathrm{previous}}}\right)$, where $r_{s_{\mathrm{previous}}}$ is the value of $r_{s}$ on the previous level and $t_0$ is the time at which the new level has been introduced. Thus the new smoothing length --\,half of the value on the previous, coarser level\,-- is reached after 1 orbit.

\subsection{Simulation sets}

We performed altogether 5 simulations, four of them are carried out with the JUPITER code, one is with the FARGOCA code. In our nominal simulations performed with both codes, the temperatures evolve freely according to the governing equations (Eqs. \ref{eq:ene} and \ref{eq:erad}) also on the planet. We refer as ``planet" the set of 32 cells around the point-mass location at [0,1,$\pi/2$] (two cells in each coordinate direction). The side of this cubic region is about 3.6 diameters of Jupiter. This is motivated by the fact that Metis, the innermost satellite of Jupiter, presently orbits at 1.8 $R_{\rm Jup}$, and that the contraction timescale of Jupiter is of the order of 1 Myr for this planetary size \citep{Guillot}. The nominal simulations resulted in very high gas temperature: $13,000$\,K at the planet location. This is probably an overestimation   because the EOS does not include dissociation and ionization; see more in section \ref{sec:discussion}. Therefore, we decided to fix the maximal temperature of the gas at the planet location (in the innermost 32 cells), by launching three other simulations with the JUPITER code with 1000\,K, 1500\,K and 2000\,K ceiling temperatures in this area. This means that at the beginning of the simulations we let the temperature on the planet evolve according to the radiative module, but when the temperature rises above the ceiling temperature, we reset it at the ceiling value, preventing it to climb further.   Even though fixing the tempearture violates energy conservation, similarly to sink-cell methods, it is still a valid detour of the problem of over-heating due to the simplistic EOS which cannot include  dissociation and ionization. Note that if a real planet was occupying these cells, it would absorb radiations from the gas, and radiate according to its own photosphere's temperature effectively acting as an energy sink.

\section{Results}

\subsection{Circumplanetary disk or circumplanetary envelope?}
\label{sec:cpe}

\begin{figure*}
\includegraphics[width=18cm]{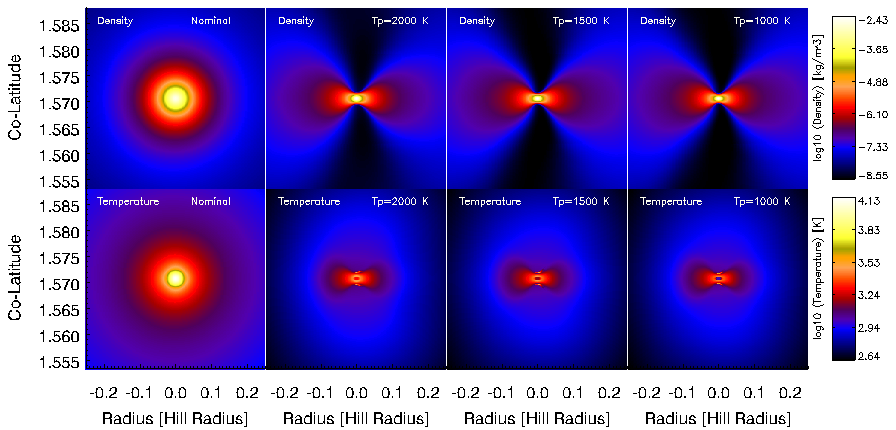}
\caption{Matrix of figures summarizing the four different simulations (Nominal, central temperature of 2000\,K, 1500\,K, 1000\,K, respectively from left to right). The first row shows the density maps, while the second shows the temperatures on a vertical slice cut through the planet along the radial direction. We can see that in the nominal simulation, where the peak temperature is over 13000\,K, a circumplanetary disk cannot form; the circumplanetary material is in a spherical envelope around the Jupiter-mass planet. Instead, when we fix the central gas temperature at the planet location to 1000\,K -- 2000\,K, always a circumplanetary disk forms. }
\label{fig:matrix}
\end{figure*}

Previous works have suggested that the formation of a circumplanetary disk is linked to the mass of the planet and the gap-opening process (e.g. \citealt{AB09a,AB09b}): small planets, which are unable to open gaps in the circumstellar disks are thought to have some circumplanetary material in the form of an envelope; instead, planets capable to carve deep enough gaps should form circumplanetary disks. However, we have found that the situation is more complex. Precisely, if the central gas temperature is high, such as in our nominal simulation, even a gap-opening, Jupiter-mass planet is forming a circumplanetary envelope instead of a disk (see Fig. \ref{fig:matrix} left column). 

In Figure \ref{fig:matrix} we show vertical slices at azimuth=0 of the density (first row) and of the temperature (second row). Each column corresponds to a different simulation: from left to right we show the nominal case, then    the fixed central temperature cases, where the gas temperature on the planet location was fixed to 2000\,K, 1500\,K, or 1000\,K. Since we have small fluctuations between the different output files, we have averaged the fields over the last 3.5 orbits of the simulations (71 outputs averaged). In the first column we clearly observe a spherical envelope. The gas temperature at the planet location is peaking at 13,000\,K. However, in all the cases with fixed central temperature a circumplanetary disk develops (columns 2 to 4). Hence, the gas temperature at the planet location seems to determine whether an envelope or a disk forms around the planet. In fact, the higher is the temperature, the more pressure supported is the disk, and the larger is its scale-height.

Furthermore, plotting the density maps on the midplane revealed that higher temperatures weaken the trace of the spiral wake in the CPD. The difference is especially striking when comparing our locally isothermal simulations in \citet{Szulagyi14} with the simulations in this work. The isothermal simulations have the lowest temperature in the circumplanetary region among all the simulations we have performed, therefore the spiral wake is the strongest. The dependence of the strength of the spiral wake on the temperature of the simulations was already pointed out in previous works e.g. by \citet{PM08} and \citet{AB09a}. 

On the temperature plots (second row on Fig. \ref{fig:matrix}) one can see that in the disk cases, there are two small regions of bright yellow color (i.e. high temperature), just above and below the central part of the CPD. We interpret these to be due to a shock between the gas infalling from the vertical direction and the disk.

\begin{figure*}
\includegraphics[width=18cm]{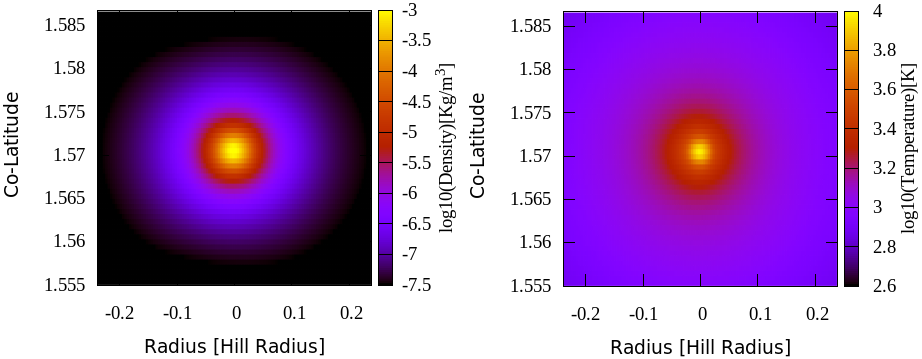}
\caption{Simulation of the nominal case made with the FARGOCA code. Left is the density map around the planet, right the temperature map. The comparison of the results of JUPITER is very good (see the first column of Fig. \ref{fig:matrix})}
\label{fig:fargoca}
\end{figure*}

We made a simulation with the FARGOCA code as well, which corresponds to the nominal simulation with JUPITER. On Figure \ref{fig:fargoca} we show the density map (left) and temperature map (right), which are quite similar to left column of Fig. \ref{fig:matrix}. The two codes with the same initial parameter file gave qualitatively similar results, namely the planet has a hot circumplanetary envelope (CPE). We recall, that the simulation made with the FARGOCA code has a resolution, which is half of the simulation made with JUPITER.

\begin{figure*}
\includegraphics[width=18cm]{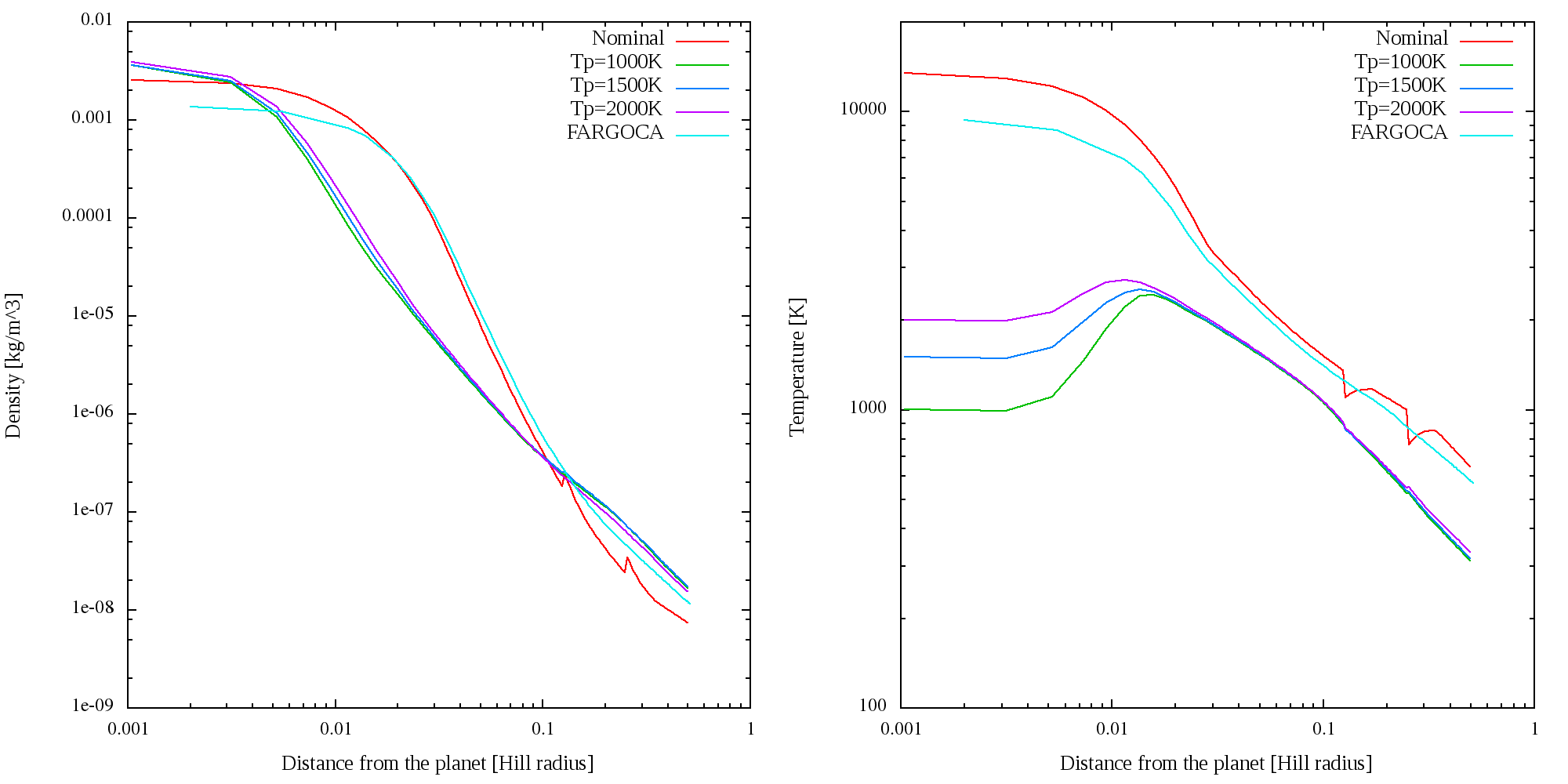}
\caption{Radial profiles of density (left) and temperature (right) at the midplane of the circumplanetary region for our four simulations with the JUPITER code, and one simulation with the FARGOCA code (as labelled). Notice that the temperatures of CPDs in the fixed central temperature simulations reach a maximum at a distance $\sim 10\% R_{\mathrm{Hill}}$.}
\label{fig:radprof}
\end{figure*}

On Figure \ref{fig:radprof}, the radial profiles of the CPD/CPE  obtained in the JUPITER simulations are compared for the densities (left) and for the temperatures (right). The density profiles are all very steep. The fixed central temperature cases almost match each other, their power law index in the outskirts of the CPD is approximatively -2.6. The nominal simulation's envelope shows a very different radial profile, which is due to geometrical reasons (the gas in the envelope has a nearly spherical symmetry). Here, the region within $0.1 R_{\mathrm{Hill}}$ contains a larger total mass than in the fixed temperature (disky) cases. However, the innermost few cells around the planet are always more massive in the fixed central temperature cases (due to the smaller temperatures, which allow a higher compression of the gas), than in the nominal simulation. The temperature profiles (right panel on Figure \ref{fig:radprof}) shows that the nominal simulation leads to temperatures higher than those in the fixed central temperature simulations in the whole domain. Even at 0.5 $R_{\mathrm{Hill}}$ away from the planet, the nominal case's envelope is still $\sim$ 400 K hotter than the fixed central temperature cases. This difference increases up to 7000 K close to the planet. In all cases the circumplanetary material is optically thick, and the temperatures in the CPD/CPE within $\sim$10-20\% of the Hill-sphere are over the dust sublimation threshold, so the opacity here is set by the gas opacities. In the fixed central temperature cases, the inner part (within a distance of 0.01-0.02 Hill radii)    of the CPD is surprisingly much hotter than the gas at the location of the planet (within 0.005 Hill radii). This is because of the     prescribed cooling with fixing the gas temperature at the planet's location. Instead, far enough from the planet this cooling effect vanishes, and the viscous heating, together with the adiabatic compressional heating can heat the CPD to be hotter than the     ceiling temperature. It is also interesting that the temperatures in the three fixed central temperature cases match beyond $0.02 R_{\mathrm{Hill}}$. The power law index of the temperature beyond this distance is around 0.6, so the disks are flared (a disk with constant aspect ratio would have a temperature proportional to $1/r$).

Comparing the nominal simulation of JUPITER and FARGOCA codes, the radial profiles are in good quantitative agreement for distances from the planet larger 
than $0.01 R_{Hill}$. This result is quite satisfactory  considering that  the two codes have  different solvers, 
different energy equations (JUPITER solves the total energy, while FARGOCA  the internal energy) and that JUPITER has proper nested meshing while FARGOCA has manual refinement.  For distances to the planet below  $0.01 R_{Hill}$ the results
are in a qualitative agreement, the quantitative difference is possibly due to
 the resolution (the potential well is more coarsely sampled in FARGOCA) or to one of the points listed above.

In summary, these findings suggest that even in the case of large mass gas-giants, the gap-opening capability does not account for whether a circumplanetary envelope or a disk forms around the planet; but the gas temperature at the planet location is the most critical factor. The characteristics of the circumplanetary material -- density and temperature profiles, rotation -- are strongly dependent on the central temperature we account for.


\subsection{Velocity and Angular Momentum}
\label{sec:velo}

\begin{figure*}
\includegraphics[width=18cm]{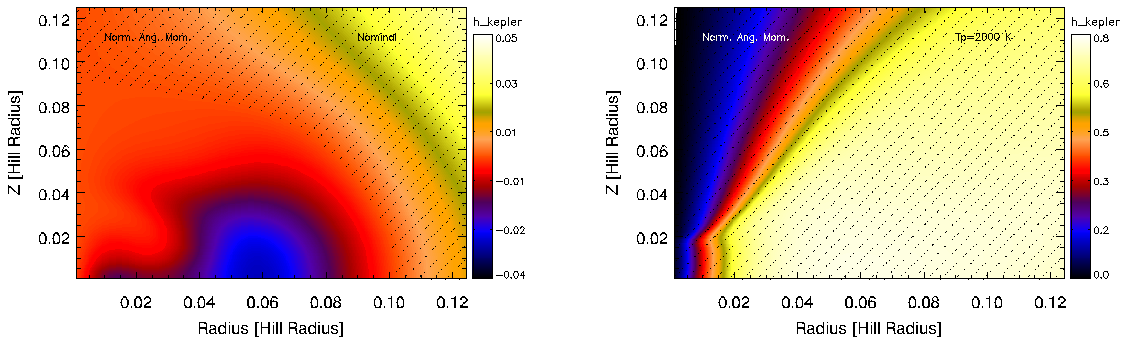}
\caption{Azimuthally averaged normalized angular momentum (z-component) from the average of 71 outputs over the last 3.5 orbits of the simulations in a non rotating frame. The normalization of the angular momentum was performed based on the local Keplerian velocity. The coordinates of the plots are cylindrical, planetocentric, so that the planet is in the left-bottom corner of each figure. The dotted areas symbolize the positive normalized angular momentum values. Left is the nominal case, with a slightly retrograde inner envelope (within 0.1 $R_{\mathrm{Hill}}$ mostly negative angular momentum), and sligthly prograde outer envelope, but overall the rotation of the envelope is almost stopped. On the right panel, a fixed central temperature ($T_p$=2000\,K) case is shown with positive normalized angular momentum values, with a maximum of 80\% Keplerian rotation.}
\label{fig:angmom}
\end{figure*}

The nominal simulation and the fixed central temperature simulations show completely different normalized angular momentum fields. The normalization of the angular momentum is based on the local Keplerian velocity, i.e. we divide the $z$-component of the angular momentum per unit mass that we measure in each cell of the simulation relative to the planet by $\sqrt{GM_p d}$ where $d$ is the radial distance of the cell from the planet in cylindrical coordinates. The angular momentum is measured in a non rotating frame centred on the planet Fig. \ref{fig:angmom} shows the above defined z-component of the normalized angular momentum for the nominal case (left) and for a fixed central temperature case ($T_p$=2000\,K case on the right) through a vertical slice. The values are azimuthally averaged (so the planet is in the left-bottom corner), and time averaged on 71 output files over the last 3.5 orbits of the simulations. The dotted areas represent the positive angular momentum values (i.e. counterclockwise rotation of the gas around the planet). One can see on the right panel that the circumplanetary disk is sub-Keplerian; it rotates with 80\% of the local Keplerian velocity (bright yellow colors). Above the disk, the gas which -- as we will see below -- falls towards the midplane, has a very low angular momentum. This means that as it hits the disk, it slows down the disk rotation. Between the three fixed central temperature cases there is not a large difference, therefore we show only one example on Fig. \ref{fig:angmom}. However, as the temperature rises, the scaleheight of the CPD is larger, and the rotation of the disk is slower (lower normalized angular momentum values).

On the left panel of Fig. \ref {fig:angmom}, the inner envelope (until $0.1\,R_{\mathrm{Hill}}$) has mostly negative normalized angular momentum. This means, that it rotates to the retrograde direction. The retrograde rotation, however, is very slow, with a maximum of 4\% of the Keplerian rotation speed. Beyond $0.1\,R_{\mathrm{Hill}}$, the envelope within the Roche-lobe is rotating prograde, but again very slowly (maximum $\sim30\%$ of Keplerian rotation). Overall, it can be said that the rotation of the envelope is almost stalled.

One could also derive the centrifugal radius $\left(R_{cent}=\frac{(J_{\mathrm{envelope}}/M_{\mathrm{envelope}})^2}{GM_p}\right)$, where $J_{\mathrm{envelope}}$ is the angular momentum of the envelope, whose mass is $M_{\mathrm{envelope}}$. This radius would correspond to the one after the envelope collapsed into a ring, while conserving its angular momentum. According to our computation $\sim 0.5 R_{\mathrm{Hill}}$ (i.e. on the last 3 refined level), this radius is one order of magnitude smaller than Jupiter's radius, therefore there would be no disk formed after this envelope has collapsed. It is important to highlight though the limitations of our simulations, e.g. the lack of a rotating planet in the middle, the over-estimated temperature which reduces rotation, etc. It is difficult to imagine that Jupiter in our Solar System went through the same phase. Our envelope possibly would not be able to produce the extended, prograde system of the Galilean moons. Our fixed temperature simulations show that there is a way to form a quasi-Keplerian disk but then, by accreting material from that disk,  Jupiter would acquire a very rapid rotation. We speculate that the temperature of Jupiter was below our gas temperature at the planet location, but larger than 2000\,K, so that it was surrounded by a prograde and sub-Keplerian, puffed-up disk.

\begin{figure*}
\includegraphics[width=18cm]{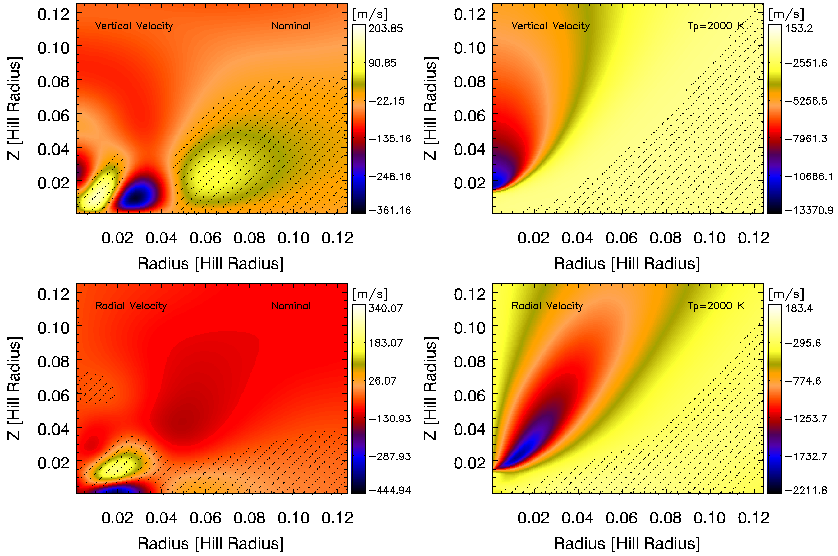}
\caption{Azimuthally averaged velocities, from the average of 71 outputs over the last 3.5 orbits of the simulations. The first row shows the vertical velocities in cylindrical, planetocentric coordinates, the second row corresponds to the radial velocities for the nominal (left column) and for the fixed central temperature ($T_p$=2000 K) simulations. The planet is the left-bottom corner in each plot. The positive velocity areas are dotted. In the nominal simulation, where an envelope formed around the Jupiter-mass planet, the sign changes in the velocities indicate a possible inner convective inner layer (until about 0.15\,$R_{\mathrm{Hill}}$), surrounded with a radiative zone. On the other hand, the fixed central temperature simulation formed a disk around the planet; here the radial velocities show clearly an inflow near the surface layer of the disk (negative $v_{rad}$), and a radial outflow near the middle of the CPD (positive $v_{rad}$, dotted area). The vertical velocities show a strong vertical influx (negative $v_{z}$), which shocks on the top of the CPD and above the planet.}
\label{fig:velo}
\end{figure*}

The velocity fields in the nominal simulation and the fixed central temperature simulations are compared in Figure \ref{fig:velo}. This figure shows the time averaged (71 outputs over 3.5 orbits), azimuthally averaged, mass-weighted, planetocentric radial velocities and vertical velocities in cylindrical coordinates for the nominal simulation (left column) and for the $T_p$=2000\,K simulation (right column). Since all the fixed temperature simulations look quite alike, we show here only the $T_p$=2000\,K case for brevity. In the nominal simulation, we see near the planet (which is placed at the left-bottom corner at 0,0 co-ordinates) alternate regions of positive and negative velocities in both the radial and vertical directions, which suggest the existence of convective motion. The convective zone is surrounded by a radiative outer layer up until the edge of the Roche-lobe. Comparing the velocity values to the fixed central temperature simulation's velocities, the difference is one-two order of magnitude. We also compared the velocity field of the FARGOCA simulation with JUPITER's. The same circulation pattern was found. 

The radial velocities of the fixed central temperature simulations (see Figure \ref{fig:velo} bottom-right insert) show a typical accretion-disk pattern: negative radial velocities on the upper layer of the CPD, so the flow is rushing toward the planet, and positive values below the upper layer (dotted area on the bottom-right panel) meaning a receding motion from the planet. The vertical velocities on the upper-right plot on Fig. \ref{fig:velo} show the strong vertical influx towards the planet, which shocks on the upper layer of the CPD above the planet. In fact, the contrast in vertical velocity between the infalling gas and the disk is about three times higher than the local sound speed. The effect of the very strong shock front was clearly visible on the temperature plots of Fig. \ref{fig:matrix}, where the shock heating highlighted this front in the upper layers of the CPD, close to the planet. In various slices on non-averaged fields we found that the vertical influx hits the shock-front so strongly, that some of it is reflected back. The angle of the vertical influx is not exactly vertical as it hits the CPD, it is slightly tilted, therefore the bounced flow escapes in the opposite direction at about the same angle. Nevertheless, most of the vertical influx will end up in the CPD and will either be accreted to the planet or leave the disk in the middle regions below the surface layer. How far from the midplane the vertical influx shocks is also determined by the local pressure, therefore the gas temperature at the planet location. Simulations with hotter central temperatures have the shock-front further away from the midplane.

As \citet{Szulagyi14} and \citet{Morbidelli14} pointed out, the vertical influx is part of a larger circulation which connects the circumstellar disk with the circumplanetary disk. Even though those simulations were isothermal, one can see the same process happening in our radiative simulations as well. As the circumstellar disk upper layers try to close the gap opened by the planet, gas enters into the gap, and free-falls onto the planet due to its gravity. This influx hits the CPD and the planet as well. In fact the CPD is mostly fed by this vertical influx as was pointed out in \citet{Szulagyi14}. Then, the gas which is not accreted onto the planet leaves the CPD in the outflow near the midplane.

\section{Conclusions \& Discussions}
\label{sec:discussion}

In this paper we have studied the circumplanetary flow around a 1 $\mathrm{M_{Jupiter}}$ planet with hydrodynamic simulations in 3D. Thanks to the nested meshes technique, we had an entire circumstellar disk in low resolution and very high resolution (80\% of the Jupiter-diameter) grid around the planet. We also implemented the energy equation and a radiative module into the JUPITER code, which accounts for both the gas and dust opacities (dust assumed to be 1\% of the gas by mass). The heating is due to viscous heating and adiabatic compression; the cooling is due to radiation and adiabatic expansion. To check our findings, we made a comparison simulation with the code FARGOCA, which also has the radiative module following the same logic, but the hydro parts are solved through different mathematical methods. 

We performed four simulations with the JUPITER code. In our nominal simulation the temperature was allowed to evolve according to the energy equations, without further constraints (resulting in a peak temperature of $\sim 13,000$\,K    at the planet location). In the other three simulations, we enforced a 1000\,K, 1500\,K and 2000\,K ceiling temperatures in the 32 cells around the point-mass planet. This change resulted in a large difference on the circumplanetary flow between the nominal and fixed temperature cases. 

While in the fixed temperature simulations a prograde rotating circumplanetary disk formed, the nominal case resulted in a very hot spherical envelope, even around this 1 $\mathrm{M_{Jupiter}}$ planet. Therefore, this finding suggests that the characteristics (temperature, mass, rotation, etc.) of circumplanetary material is mostly determined by the   gas temperature at the planet location rather than the planetary mass. Moreover, the ability to form a circumplanetary disk does not depend simply on the ability of the planet to open a gap in the circumstellar disk, since even a gap-opening 1 $\mathrm{M_{Jupiter}}$ planet can form a circumplanetary envelope, like the low-mass planets, if the central gas temperatures are very high.

We overall found that higher temperatures reduce the disk's rotation and weaken the trace of the spiral wake in the subdisk,   that can alter accretion, in accordance with previous works (e.g.\citealt{AB09a}, \citealt{PM08}). In all of our simulations the circumplanetary material is optically thick with very steep temperature and density profiles. Moreover, the nominal case with the circumplanetary envelope has an internal convection layer of 0.15\,$R_{\mathrm{Hill}}$, surrounded by a radiative layer extended up to the edge Roche-lobe. This envelope has   a very limited rotation. In the fixed central temperature simulations, however the subdisk shows moderately sub-keplerian, prograde rotation and it is fed by a strong, vertical influx arising from the top layers of the circumstellar disk and the walls of the gap, which then shocks on the CPD surface.

The used EOS in this work, however, overestimates temperatures by neglecting dissociation and ionization of hydrogen. In order to estimate the magnitude of this effect, we compared the changes in specific entropy at the location of the planet in our nominal simulation to those recalculated from our pressures and temperatures using a realistic EOS \citep{S95} that accounts for dissociation and ionization. From this test we could conclude that the temperatures are indeed overestimated, and that the specific entropies are high enough to be consistent with the ``hot start scenario" of planet formation (e.g., \citealt{MC14}). However, dedicated simulations are needed to investigate this in detail which will be part of a future publication. 


\section*{Acknowledgments}

The authors are thankful for the anonymous referee for his/her comments. J. Szul\'agyi acknowledges the support from the Capital Fund Management's J.P. Aguilar PhD fellowship and the ETH Post-doctoral Fellowship from the Swiss Federal Institute of Technology (ETH Zurich). The Nice group is thankful to the Agence Nationale pour la Recherche under grant ANR-13-BS05-0003-01 (MOJO). J. Sz. and F. M. acknowledges the support from Universidad Nacional Aut\'onoma de M\'exico grant PAPIIT IA101113, F. M. also from CONACyT’s grant 178377. This work has been in part carried out within the frame of the National Centre for Competence in  Research  ``PlanetS"  supported by  the  Swiss  National Science Foundation. Computations have been done on the ``Mesocentre SIGAMM" machine, hosted by Observatoire de la C\^ote d’Azur. Part of this work was performed using HPC resources from GENCI [IDRIS] (Grant 2015, [100507]).

\label{lastpage}

\end{document}